\documentclass{edp-conf}
\usepackage{graphicx}
\usepackage{amssymb} 
\begin{document}

\TitreGlobal{SF2A 2004}

\title{Excitation rates of p~modes: mass luminosity relation across the  HR diagram
}
\author{Samadi, R.}\address{Observatoire de Paris, Meudon, France}
\author{Georgobiani, D.}\address{Center for Turbulence Research, Stanford University NASA Ames Research Center, Moffett Field, USA}
\author{Trampedach, R.}\address{Research School of Astronomy and Astrophysics,
   Mt.\ Stromlo Observatory, Cotter Road, Weston ACT 2611, Australia}
\author{Goupil, M.J.$^1$}
\author{Stein,~R.F.}\address{Department of Physics and Astronomy, Michigan State University, Lansing, USA}
\author{Nordlund, {\AA}.}\address{Niels Bohr Institute for Astronomy Physics and Geophysics, Copenhagen, Denmark}

\runningtitle{Modeling the p modes excitation with 3D simulations}
\setcounter{page}{237}
\index{Samadi, R.}
\index{Georgobiani, D.}
\index{Trampedach, R.}
\index{Goupil, M.J.}
\index{Stein, R.F.}
\index{Nordlund, A.}


\maketitle
\begin{abstract} 
We compute the rates $P$ at which  energy is injected into the p modes for a set of 3D simulations of  outer layers of stars.
 We found that $P_{\rm max}$ -~ the maximum in $P$ ~ - scales as $(L/M)^{s}$ where $s$ is the slope of the power law,  L and M are the luminosity and the mass of the 1D stellar models associated with the simulations. The slope is found to depend significantly on the adopted representation for the turbulent eddy-time correlation function, $\chi_k$. According to the expected performances of COROT, it will likely be possible to measure $P_{\rm max}$  as a function of $L/M$ and to constrain the properties of stellar turbulence as the turbulent eddy time-correlation.
\end{abstract}
%
\section{Introduction}

Stars with masses $  M \lesssim 2 M_\odot$ have upper convective zones where the stochastic excitation of p~modes takes place. Models of stochastic excitation yield the rate $P$ at which p modes are excited by turbulent convection but require an accurate knowledge of the time averaged and -~ above all ~ - the dynamic properties of turbulent convection. The latter are represented in the approach of Samadi \& Goupil (2001) by $\chi_k$, the frequency component of the kinetic energy spectrum, also designed as the convective eddy time-correlations.  A Gaussian   is usually assumed for $\chi_k$ , {\it i.e.}  in other word it is usually assumed that two distant points in the turbulent medium are uncorrelated.  Here, we investigate two different forms for $\chi_k$:  the Gaussian and the Lorentzian. 

We consider a set of 3D simulations of solar-like oscillating stars.
We compute internal structure of 1D stellar models and associated eigenfunctions consistent with the 3D simulations.
The   3D simulations provide constraints on  the turbulent kinetic spectrum $E(k)$. Theses constraints enable us to compute $P_{\rm max}$, the maximum in $P$, for the 1D internal models consistent with the 3D simulations (see details in Samadi et al, 2004).

\section{Results and conclusion}

\begin{figure}[h]
   \centering
   \includegraphics[width=9cm]{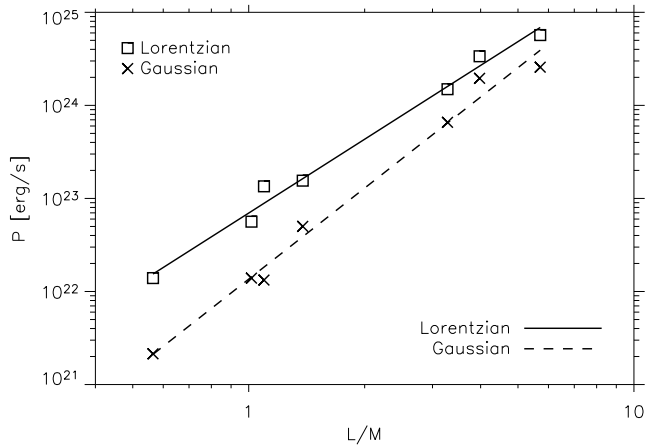}
\vspace{-0.5cm}
      \caption{Computed ${P}_{\rm max}$ versus $L/M$:
The crosses assume a Gaussian for  $\chi_k$ and the squares a Lorentzian respectively.
The lines are the results of fitting each set of dots with a power law of the
form $(L/M)^{s}$ where $s$ is the slope of the power law. }
       \label{figure_mafig}
   \end{figure}

Results of the calculations of ${P}_{\rm max}$  are shown in Fig.~(1).
We find that $P_{\rm max}$  scales as $(L/M)^s$ where $s$  is the slope of the scaling law, $L$ is the   luminosity and $M$ is the mass of the 1D~models associated with the 3D simulations. The value found for $s$ is  sensitive to the choice for $\chi_k$: Indeed the Gaussian results in  $s=3.3$  and the Lorentzian in $s=2.6$.
Solar seismic observations can clearly distinguish between the two forms for $\chi_k$: they  favor the Lorentzian (see Samadi et al, 2003).
For other stars for which stochastically excited p modes are expected, the Lorentzian is likely to remain the correct form for  $\chi_k$, but this needs to be confirmed. 
The high-quality data of the COROT instrument will enable us to derive $P_{\rm max}$ as a function of L and M for the different target stars.
These data will very likely enable us to  determine the appropriate model for the eddy time-correlations  ($\chi_k$).


\end{document}